\documentclass[aps,prl,preprint,superscriptaddress,floatfix]{revtex4-1}
\usepackage{amsmath}
\usepackage{amsfonts} 
\usepackage{graphicx}
\usepackage{verbatim}
\usepackage{natbib}
\usepackage{threeparttable}
\usepackage{xcolor}
                                          
\setcitestyle{super}

\begin{document}

\begin{center}
\textbf{\large Spectroscopic and Interferometric Sum-Frequency Imaging of Strongly Coupled Phonon Polaritons in SiC Metasurfaces }\\[0.5em]
%\textbf{\large Imaging Propagating Phonon Polaritons in SiC Metasurfaces with Sum-Frequency Spectro-Microscopy}\\[0.5em]

Richarda Niemann,$^{1,*}$ Niclas S. Mueller,$^{1,*}$ Sören Wasserroth,$^{1}$ Guanyu Lu,$^{2,\parallel}$ Martin Wolf,$^{1}$ Joshua D. Caldwell,$^{2,3}$ and Alexander Paarmann$^{1, \dagger}$\\[0.2em]

\begin{small}
\textit{$^1$ Fritz-Haber-Institut der Max-Planck-Gesellschaft, Faradayweg 4-6, 14195 Berlin, Germany\\}

\textit{$^2$ Department of Mechanical Engineering, Vanderbilt University, Nashville, TN 37235, USA}

\textit{$^3$ Interdisciplinary Materials Science Graduate Program, Vanderbilt University, Nashville, TN 37235, USA\\}
\textit{$^\parallel$ Present address: Department of Chemical and Biological Engineering, Northwestern University, 2145 Sheridan Road, Evanston, IL 60208, USA\\}
\vspace{0.5 cm}
\textit{$^*$ These authors contributed equally to this work.\\}
\textit{$^\dagger$ correspondence to: alexander.paarmann@fhi-berlin.mpg.de}

\end{small}

\date{\today}
    
\end{center}

Phonon polaritons enable waveguiding and localization of infrared light with extreme confinement and low losses. The spatial propagation and spectral resonances of such polaritons are usually probed with complementary techniques such as near-field optical microscopy and far-field reflection spectroscopy. Here, we introduce infrared-visible sum-frequency spectro-microscopy as a tool for spectroscopic imaging of phonon polaritons. The technique simultaneously provides sub-wavelength spatial resolution and highly-resolved spectral resonance information. This is implemented by resonantly exciting polaritons using a tunable infrared laser and wide-field microscopic detection of the upconverted light. We employ this technique to image hybridization and strong coupling of localized and propagating surface phonon polaritons in metasurfaces of SiC micropillars. Spectro-microscopy allows us to measure the polariton dispersion simultaneously in momentum space by angle-dependent resonance imaging, and in real space by polariton interferometry. Notably, we directly visualize how strong coupling affects the spatial localization of polaritons, inaccessible with conventional spectroscopic techniques. We further observe the formation of edge states at excitation frequencies where strong coupling prevents polariton propagation into the metasurface. Our approach is applicable to the wide range of polaritonic materials with broken inversion symmetry and can be used as a fast and non-perturbative tool to image polariton hybridization and propagation.

\newpage
\section{Introduction}
Phonon polaritons are mixed light-matter states that emerge from the strong coupling of photons with optical phonons in polar semiconductors. Propagating phonon polaritons at the surface (SPhP) and in the bulk of these materials enable waveguiding and localization of infrared (IR) light with extreme confinement and low losses.\cite{caldwell2015low,Foteinopoulou2019, Zhang2021, Wu2022, Gubbin2022} The anisotropic crystal structure of many polar materials leads to exotic phenomena, such as hyperbolic dispersion, highly directional propagation, hyperlensing, and negative refraction.\cite{Zhang2021, Wu2022, He2022, Galiffi2023} Additional functionality is achieved by sub-wavelength structuring of polaritonic materials\cite{li2018infrared, Wu2022}, e.g.\ by building periodic lattices of subwavelength meta-atoms and molecules that support localized surface phonon polaritons (LSPhPs).\cite{Caldwell2013,razdolski2016resonant} The hybridization of localized and propagating polaritons in such metasurfaces enables strong coupling,\cite{Gubbin2016,razdolski2018second,Lu2021,Hu2022} sensing of molecules,\cite{Bylinkin2021,autore2018boron} directional thermal emission,\cite{Greffet2002,Lu2021,Lu2022_SubArrays} or topologically protected edge states.\cite{Guddala2021} The properties of polaritonic metasurfaces strongly depend on the excitation wavelength because of the strong material dispersion induced by the underlying phonon resonances. For a full experimental characterization one therefore needs tools that combine spectral information across a broad range of frequencies with spatial information from sub-diffractional to mesoscopic length scales. 

The spatial propagation characteristics and spectral resonances of polaritons are typically measured with complimentary techniques. A powerful tool to image polariton propagation is scattering-type scanning near-field optical microscopy (s-SNOM), which provides deep sub-wavelength spatial information through interferometric detection of the scattered polaritonic near fields by a metallic tip.\cite{Huber2005,Folland2019} Attaching an s-SNOM to a tunable IR laser in principle enables combined spectral and spatial information through multi-spectral imaging.\cite{alfaro2021hyperspectral, Kusch2021} This is however limited to a few selected wavelengths because of the long image acquisition times in scanning-probe imaging. Detailed spectral information can instead be obtained at selected positions with nano-Fourier transform infrared (nano-FTIR) spectroscopy by using a broadband light source.\cite{Amarie:09} This is complimented by far-field techniques such as reflection spectroscopy,\cite{Caldwell2013} prism coupling,\cite{passler2018strong,runnerstrom2018polaritonic,ni2023observation,Passler2022} or thermal emission spectroscopy\cite{Greffet2002,Lu2021, LuAPL2021} that provide well resolved spectral information while spatially averaging over large sample areas. With these currently established techniques it is especially challenging to characterize the resonant properties of polaritonic metasurfaces in which polaritons propagate over $>$100\,µm distances along corrugated surfaces while near fields are locally confined to sub-µm volumes. 

Here, we introduce sum-frequency spectro-microscopy as a far-field optical tool to image polaritons in metasurfaces with highly-resolved spatial and spectral information. A tunable infrared free-electron laser (IR-FEL) excites polaritons across the entire field-of-view and a visible (VIS) laser upconverts the local polaritonic near fields through sum-frequency generation (SFG).\cite{Kiessling2019, Niemann2022} The SFG signal from the metasurface is imaged in a wide-field optical microscope with a spatial resolution that is ultimately set by the diffraction limit of the visible SFG wavelength. By scanning the IR wavelength we obtain highly-resolved spatio-spectral information, much faster than in scanning-focus approaches. We use this approach to image a SiC micropillar metasurface in which localized and propagating surface polaritons hybridize through strong coupling.\cite{Gubbin2016,Lu2021,Hu2022} SFG spectro-microscopy allows us to simultaneously measure the polariton dispersion using resonant spectral imaging as in prism coupling\cite{passler2018strong,Passler2022,ni2023observation} and thermal emission spectroscopy\cite{Greffet2002,Lu2021} and spatial interferometry as in s-SNOM.\cite{Huber2005,barnett2022investigation} Remarkably, when tuning the IR wavelength across the strong coupling dispersion, we find a continuous change of the polariton propagation length directly reporting on the transition from propagating to localized polaritons that scales with the polariton mixing fraction. We furthermore observe a localization of polaritons at the metasurface edges within the strong coupling bandgap where propagation across the metasurface is forbidden. Our fast and non-perturbative far-field experimental approach can be applied to the wide range of metasurfaces made from inversion-broken polaritonic materials.

\begin{comment}
\subsection{Paragraph 1}
\begin{itemize}
    \item Propagating SPhPs provide spatial coherene, while LSPhPs spatial confinement and small mode volumes
    \item Hybridization of both combines their advantages
    \item While SPhPs can propagate over $>$100um distances, LSPhPs confine light to sub µm volumes
    \item SPhPs can be furthermore launched by edges, defects, and optical antennas, and several SPhPs can interfer with each other, leading to complex propagation patterns
    \item Characterizing hybridized (propagating and localized) polaritons therefore requires imaging of large areas, and at the same time to resolve sub-wavelength features of the LSPhPs
\end{itemize}

\subsection{Paragraph 2}

Previous methods to study polaritons in SiC arrays:
\begin{itemize}
    \item Reflection (poor spatial resolution)
    \item Thermal emission (poor spatial resolution)
    \item SNOM (cannot scan large areas with high resolution if sample not flat, hard to do spectroscopy)
    \item A powerful complementary tool is sum frequency spectro-microscopy (refer to APL paper)
\end{itemize}

Open questions to motivate:
\begin{itemize}
    \item how (and how long) do polaritons propagate in SiC pillar lattices?
    \item does strong coupling prevent propagation along edges as well?
    \item can polaritons propagate beyond the SiC array?
\end{itemize}
\end{comment}

\section{Results}
We implement a wide-field SFG microscope to image the tightly confined electric fields of localized and propagating surface phonon polaritons (Fig.\,1a,b, Methods). A tunable free-electron laser (FEL)\cite{ Schoellkopf2015} is used to resonantly excite polaritons with coherent infrared pulses ($\omega_\mathrm{IR} = 750 – 1050\,\mathrm{cm}^{-1}$, duration $\approx$\,5\,ps, linewidth $\approx 0.6\%$), while a synchronized visible laser illuminates the sample from the backside ($\lambda_\mathrm{VIS} = 532$\,nm, duration $\approx$\,12\,ps).\cite{Niemann2022} The nonlinear mixing of the two lasers generates an SFG signal that scales with the local electric field intensity, resonantly enhanced by the phonon polaritons on the sample surface.\cite{Liu2008,Kiessling2019} For materials with broken inversion symmetry such as silicon carbide (SiC), the SFG signal is generated from the bulk of the crystal, i.e., from the polariton mode volume.\cite{Kiessling2019,Niemann2022} The upconverted SFG light is collected with a high numerical aperture objective (50$\times$, 0.55NA) and detected by a gated CCD camera, exploiting the very precise optics and sensitive detectors in the visible spectral range for nonlinear IR spectro-microscopy. Both, the infrared and the visible laser illuminate the entire field of view of the optical microscope which allows wide-field imaging of a large $275\times275$\,µm$^2$ area of the sample. The spatial resolution is limited here by the visible SFG wavelength to $\approx 1.4$\,µm ($\lambda_{IR}/9$), which is much shorter than the IR wavelength, making it possible to resolve polariton near fields.\cite{Niemann2022} Compared to scanning-focus microscopes\cite{Kiessling2019} our wide-field implementation avoids laser damage and allows to record SFG images with an acquisition time of 10\,s/image, at the expense of requiring a high-power infrared laser.\cite{Hanninen2017, Shah2020} The tunable infrared laser further enables spectroscopic imaging by step-wise scanning the infrared wavelength across polariton resonances and recording SFG images (Fig.\,1b). Our technique thereby provides combined spatial and spectral information about polaritons.

\begin{figure*}[h!]
    \includegraphics[width=16.5cm]{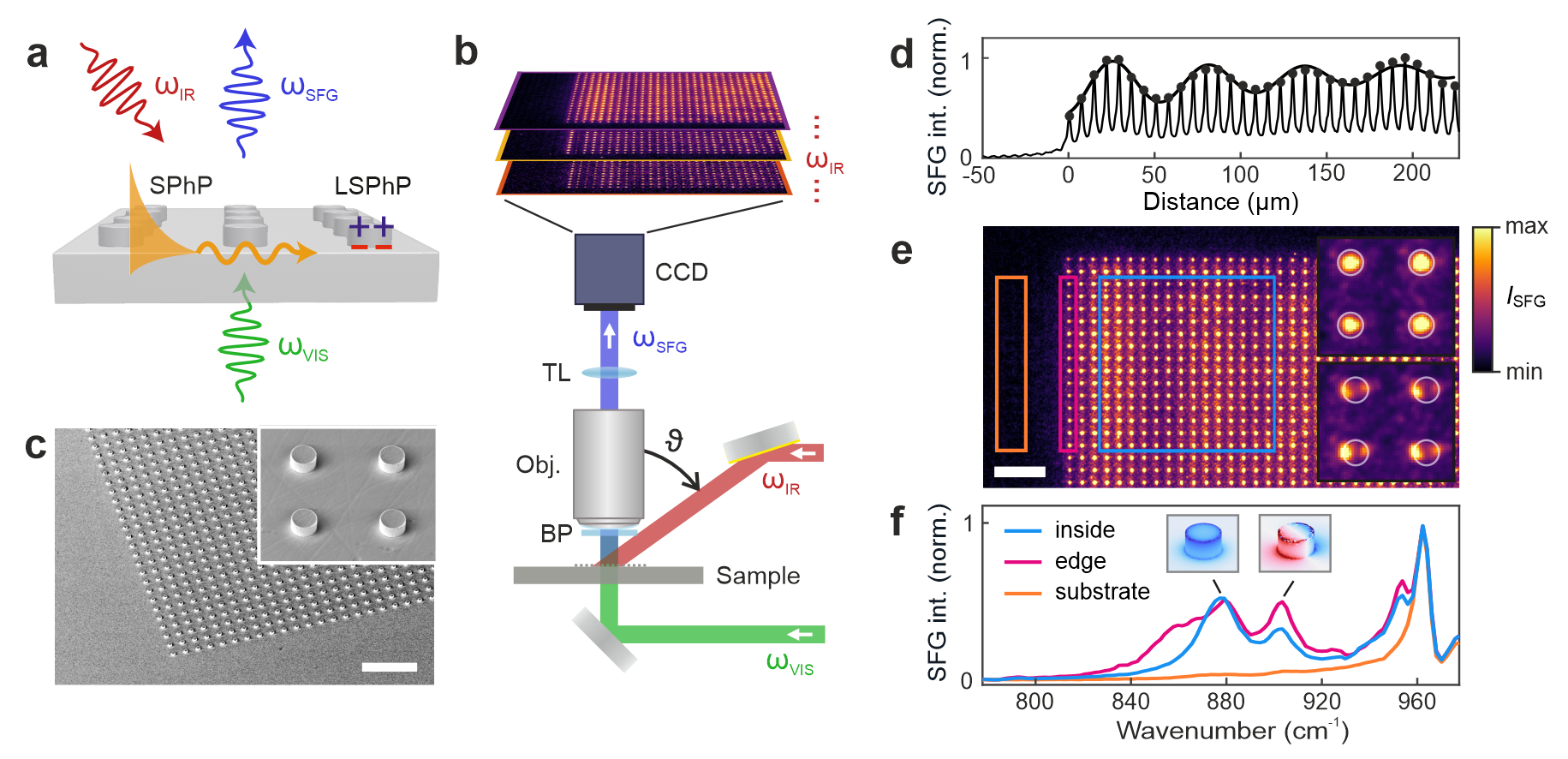}
    \caption{Sum-frequency spectro-microscopy of phonon polaritons in SiC micropillar arrays. (a) Sum-frequency generation ($\omega_\mathrm{SFG}$) with a mid-infrared ($\omega_\mathrm{IR}$) and visible ($\omega_\mathrm{vis}$) laser of propagating surface phonon polaritons (SPhPs) and localized SPhPs (LSPhPs) in a SiC micropillar array. (b) Wide-field SFG microscope using side illumination with a tunable free-electron IR laser (red) and illumination through the sample with a visible laser (green). The SFG light (blue) is detected with a microscope objective (Obj.) and the visible laser is blocked with a band pass filter (BP). A tube lens (TL) and CCD detector are used for hyperspectral SFG imaging at different IR frequencies. (c) Scanning electron microscopy images of SiC micropillar array. 30\,µm scale bar, 13.9\,µm edge length of inset.  (d) Horizontal line cut of SFG intensity in (e) from full vertical binning. (e) SFG microscopy image of SiC micropillar array at $\omega_\mathrm{IR} = 886\ \mathrm{cm}^{-1}$. Scale bar is 30\,µm. Insets are SFG images of monopole mode (top, $\omega_\mathrm{IR} = 876\ \mathrm{cm}^{-1}$) and dipole mode (bottom, $\omega_\mathrm{IR} = 902\ \mathrm{cm}^{-1}$) of SiC pillars. White lines indicate size of SiC pillars (2.5\,µm diameter); image sizes are 14$\times$14\,µm. (f) SFG spectra from average intensity of entire metasurface (inside, blue), edge of metasurface (magenta), and SiC substrate beside metasurface (orange). Insets show monopole (left) and dipole (right) LSPhP resonances of micropillars through their simulated electric field normal to the surface. Areas in which spectra were integrated are indicated in (e) with corresponding colors.}
    \label{fig:Fig1}
\end{figure*}

We apply the SFG microscope to image phonon polaritons in mm-sized arrays of 4H-SiC micropillars (Fig.\,1a, c).\cite{Gubbin2016, razdolski2018second, Lu2021, Hu2022} These metasurfaces consist of lithographically fabricated SiC disks on top of a SiC substrate with a sub-wavelength periodicity $P = 7.3$\,µm. The metasurface supports both propagating surface phonon polaritons (SPhPs) and localized SPhPs (LSPhPs) in the spectral range $\omega_\mathrm{IR} = 800 - 960\,\mathrm{cm}^{-1}$ of the SiC Reststrahlen band. The polaritons can further hybridize through strong coupling which modifies their dispersion and propagation, as will be discussed below. This has been characterized previously with angular reflection and thermal emission spectroscopy, as well as s-SNOM, which are techniques that either provide spectral or spatial information.\cite{Gubbin2016, Lu2021, Lu2022_SubArrays, Hu2022} In an SFG microscope image of the metasurface each individual micropillar appears as a bright spot from the excitation of LSPhPs (Fig.\,1e). A dimmer SFG signal around the pillars as well as vertical interference fringes further hint to a contribution from propagating polaritons (Fig.\,1d). Typically, much reduced SFG signal is observed outside the metasurface due to the lack of  polaritonic field enhancement on the flat SiC surface in the Reststrahlen band spectral range.\cite{Kiessling2019}

We use spectroscopic imaging to measure the spectral response of different positions and areas of the metasurface (colored boxes in Fig.\,1e) by scanning $\omega_\mathrm{IR}$ in steps of 2\,cm$^{-1}$, resulting in the local spectra shown in Fig.\,1f. The spectrum from the integrated SFG intensity of the entire metasurface (blue) is dominated by the LSPhP resonances of the individual micropillars (see insets). At $\omega_\mathrm{M} = 864\,\mathrm{cm}^{-1}$ the out-of-plane component of the infrared laser field excites a monopole resonance that corresponds to a vertical charge separation in the microdisks.\cite{Caldwell2013, Gubbin2022} At $\omega_\mathrm{D} = 906\,\mathrm{cm}^{-1}$, instead, the in-plane polarization component excites a dipole mode with lateral charge separation, which leads to two intensity maxima at the edges of the disks. The resonances in Fig.\,1f are shifted from the uncoupled states because of polariton hybridization, as will be discussed below. These near-field distributions of the monopolar and dipolar modes can directly be resolved in the SFG microscope images (Figs.\,1e, insets), which demonstrates the sub-wavelength spatial resolution of our technique.\cite{Niemann2022} In addition to the LSPhPs, a pronounced resonance occurs at $\omega_\mathrm{LO} = 962\,\mathrm{cm}^{-1}$, which arises at the longitudinal-optical (LO) phonon frequency of the SiC substrate.\cite{Paarmann2016} This resonance is associated with an enhancement of the local optical fields at the SiC-air interface at the LO frequency,\cite{Kiessling2019} and thus is also detected outside the metasurface (Fig.\,1e,f, orange). Interestingly, the SFG spectrum of the metasurface edge differs profoundly from the spectrum recorded inside the metasurface with an additional resonance emerging at $\omega_\mathrm{edge} \approx 860\,\mathrm{cm}^{-1}$ (Fig.\,1e,f, magenta) that cannot be assigned to the localized resonances of the individual pillars observed inside the metasurface (Fig.\,1e,f, blue). We will analyze this in detail below.

The coupling of propagating SPhPs at the SiC surface and the localized monopole and dipole modes of the micropillars leads to hybridized polaritons (Fig.\,2a,b).\cite{Gubbin2016, Lu2021, Hu2022} This is visible through pronounced changes of the SPhP dispersion, obtained here from the simulated angular absorption of the metasurface (Fig.\,2b, Methods). The polariton dispersion $\omega_\mathrm{P}$ shows two anti-crossings with the $\omega_\mathrm{M}$ and $\omega_\mathrm{D}$ resonances of the micropillars, forming three hybridized polariton branches with different energies. We analyze this hybridization with a three-mode coupled oscillator model with a coupling matrix\cite{Lu2021}
\begin{equation}
\label{eq:CouplingMatrix}
\mathcal{M} =
\begin{pmatrix}
\omega_\mathrm{SPhP} & g_1 & g_2\\
g_1 & \omega_\mathrm{M} & 0\\
g_2 & 0 & \omega_\mathrm{D}\\
\end{pmatrix},
\end{equation}
where $g_1$ describes the SPhP-monopole, and $g_2$ the SPhP-dipole coupling. We neglect direct coupling between the monopole and dipole resonances because of their negligible spectral and spatial overlap. The eigenvalues of the coupling matrix give the three hybridized polariton branches. The model excellently matches the simulated dispersion when assuming coupling strengths $g_1 = 22\,\mathrm{cm}^{-1}$ and $g_2 = 5\,\mathrm{cm}^{-1}$. A clear spectral splitting is visible at both anti-crossings indicating strong coupling of both the monopole and the dipole mode with the SPhP. This is in good agreement with previous experiments using angle-resolved reflection and thermal emission spectroscopy.\cite{Gubbin2016, Lu2021} The SPhP-monopole coupling is much larger than the SPhP-dipole coupling because the SPhP near fields match the out-of-plane polarization of the monopole mode, while the dipole mode fields are mostly in-plane. 

\begin{figure*}[h!]
    \includegraphics[width=16.6cm]{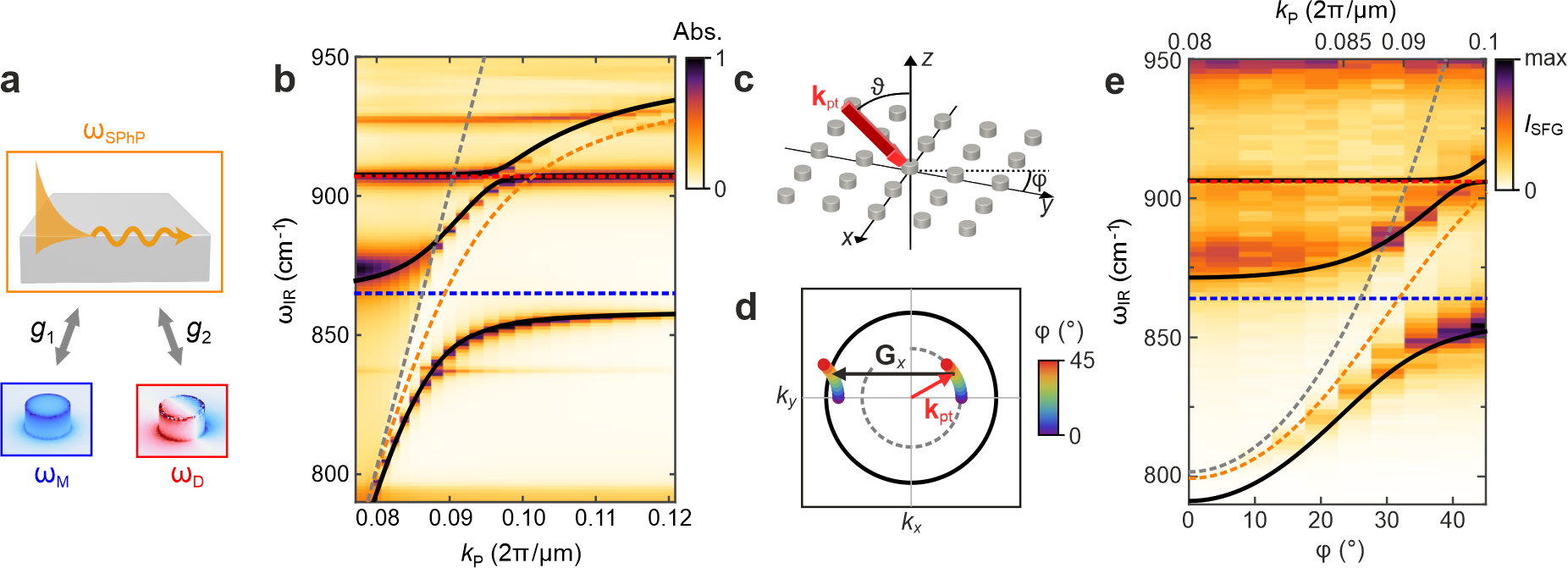}
    \caption{Dispersion of hybridized phonon polaritons from resonant spectral imaging. (a) Coupling of propagating surface phonon polariton ($\omega_\mathrm{SPhP}$) and localized monopole $\omega_\mathrm{M}$ and dipole $\omega_\mathrm{D}$ modes of SiC pillars. (b) Calculated dispersion of hybridized phonon polaritons (black lines) with dashed lines showing uncoupled states $\omega_\mathrm{SPhP}$ (orange), $\omega_\mathrm{M}$ (blue), $\omega_\mathrm{D}$ (red), and free space photons (grey). Contour plot shows optical absorption from full-wave simulation. Parameters: periodicity $P=7.27$\,µm, pillar diameter 2.5\,µm, and height 1\,µm. (c) Sketch of incidence angle $\vartheta$ of the IR laser and azimuthal angle $\varphi$ in experiments. (d) Isofrequency contour of hybridized surface polariton (black) and in-plane projection of photon (grey dashed line). A polariton can be excited through momentum matching of the incident light $\mathbf{k}_\mathrm{pt}$ with a reciprocal lattice vector $\mathbf{G}_x$ for a specific incidence angle $\varphi$. (e) Polariton dispersion measured with SFG spectro-microscopy by scanning the azimuthal angle $\varphi$. Lines show prediction by analytic model with same parameters as in (b), and $\vartheta = 46^\circ$.}
    \label{fig:Fig2}
\end{figure*}

SFG spectro-microscopy gives direct access to the hybridized polariton dispersion by following the polariton resonances in the  SFG spectra for different azimuthal angles $\varphi$ of the metasurface with respect to $\omega_\mathrm{IR}$ (Fig.\,2c-e). The dispersion of the hybridized polaritons is mostly below the light line, which in general prevents direct excitation with far-field radiation because of the much larger polariton momentum (compare black and grey lines in Fig.\,2b,d). Momentum matching is however possible here through the periodic lattice of the metasurface, which provides reciprocal lattice vectors $\mathbf{G}_x$ and $\mathbf{G}_y$ for far-field excitation\cite{Gubbin2016}
\begin{equation}
    \mathbf{k}_\mathrm{exc} = \mathbf{k}_{\mathrm{pt}, \mathrm{ip}} + m \mathbf{G}_x + n \mathbf{G}_y,
\end{equation}
with $m, n$ integers, and $\mathbf{k}_{\mathrm{pt}, \mathrm{ip}}$ the in-plane photon momentum. A polariton with momentum $k_P$ is excited for a combination of the azimuthal angle $\varphi$ of the sample and the incidence angle $\vartheta$ of the IR laser that fulfill
\begin{equation}
    k_\mathrm{P}^2 = \left( k_\mathrm{pt} \sin{\vartheta}\cos{\varphi} + m \frac{2\pi}{P} \right)^2 + \left( k_\mathrm{pt} \sin{\vartheta}\sin{\varphi} + n \frac{2\pi}{P} \right)^2,
\end{equation}
where $P$ is the lattice periodicity. This is illustrated in Fig.\,2d for an isofrequency contour of the lower polariton branch at fixed $\vartheta = 46^\circ$ as in experiments. At a rotation angle $\varphi=30^\circ$ momentum matching is possible with $m=-1$ and $n=0$. More generally, a rotation of the metasurface from $\varphi=0-45^\circ$ gives access to a range of polariton momenta for different IR frequencies. We use this dependence to map out the dispersion of the hybridized polaritons by recording the local SFG spectra for different rotation angles $\varphi$ (Fig.\,2e). The measured dispersion excellently matches the analytic model and simulations. Compared to simulations, the spectral width of the polaritonic resonances is larger, which could result from disorder in the lattice, and excitation with a narrow distribution of $\vartheta$ and $\varphi$ because of a mild focusing of the IR laser beam, as well as its finite bandwidth. While the anti-crossing of $\omega_\mathrm{D}$ and $\omega_\mathrm{SPhP}$ can no longer be resolved, the coupling of $\omega_\mathrm{M}$ and $\omega_\mathrm{SPhP}$ is clearly within the strong coupling regime with $g_1 = 24\,\mathrm{cm}^{-1} > (\gamma_\mathrm{M} + \gamma_\mathrm{SPhP})/2 = (9.0 + 7.9)/2\,\mathrm{cm}^{-1}=8.5\,\mathrm{cm}^{-1}$, where the line widths $\gamma_\mathrm{M}$, $\gamma_\mathrm{SPhP}$ of the uncoupled states were previously measured with thermal emission spectroscopy.\cite{Lu2021}  

In addition to the direct launching of polaritons via the metasurface's lattice vectors, we also observe interference fringes of the SFG intensity as could be expected from propagating polaritons within the metasurface (Fig.\,3a, b). The fringes occur predominantly along the perpendicular direction to the incidence of the IR laser and parallel to the metasurface edge, with a decreasing amplitude away from the edge (see also SI Section S1 for opposite edge). This observation suggests that polaritons are launched by the edge of the metasurface. The fringe period and propagation length strongly depends on the IR laser frequency and occurs independent of the momentum matching conditions for far-field excitation (i.e.\ at $\varphi=0$). We analyze the fringes with a model assuming the interference of an edge-launched polariton with the incident IR laser field (Fig.\,3c), similar to the interpretation of interference fringes in s-SNOM images.\cite{barnett2022investigation} The spatial modulation of the SFG intensity within the metasurface is well fit with (Fig.\,3b, thick lines)
    \begin{equation}
    \label{eq:InterferometricImaging}
        I(x)=\vert A e^{i k_\mathrm{P} x + i \xi} + B e^{i k_\mathrm{pt} \sin{\vartheta} x} \vert^2 + I_0 + I_1 x,
    \end{equation}
with the complex polariton momentum $k_\mathrm{P}$, a phase shift $\xi$, amplitudes $A$ and $B$ of the polariton and photon waves, and a linear background $I_0 + I_1 x$. The fringe period $\Lambda_x = 2\pi/(\mathrm{Re}[k_\mathrm{P}] - k_\mathrm{pt} \sin{\vartheta}$) is much larger than the actual polariton wavelength $\lambda_\mathrm{P}$, consistent with polaritons co-propagating with the light field\cite{barnett2022investigation} launched by the vertical edge. Launching by the top edge, perpendicular to $\mathbf{k}_\mathrm{pt}$, would result in a fringe period $\Lambda_y = \lambda_\mathrm{P}$, but apparently is far less efficient and thus not observed.

\begin{figure*}[h!]
    \includegraphics[width=16cm]{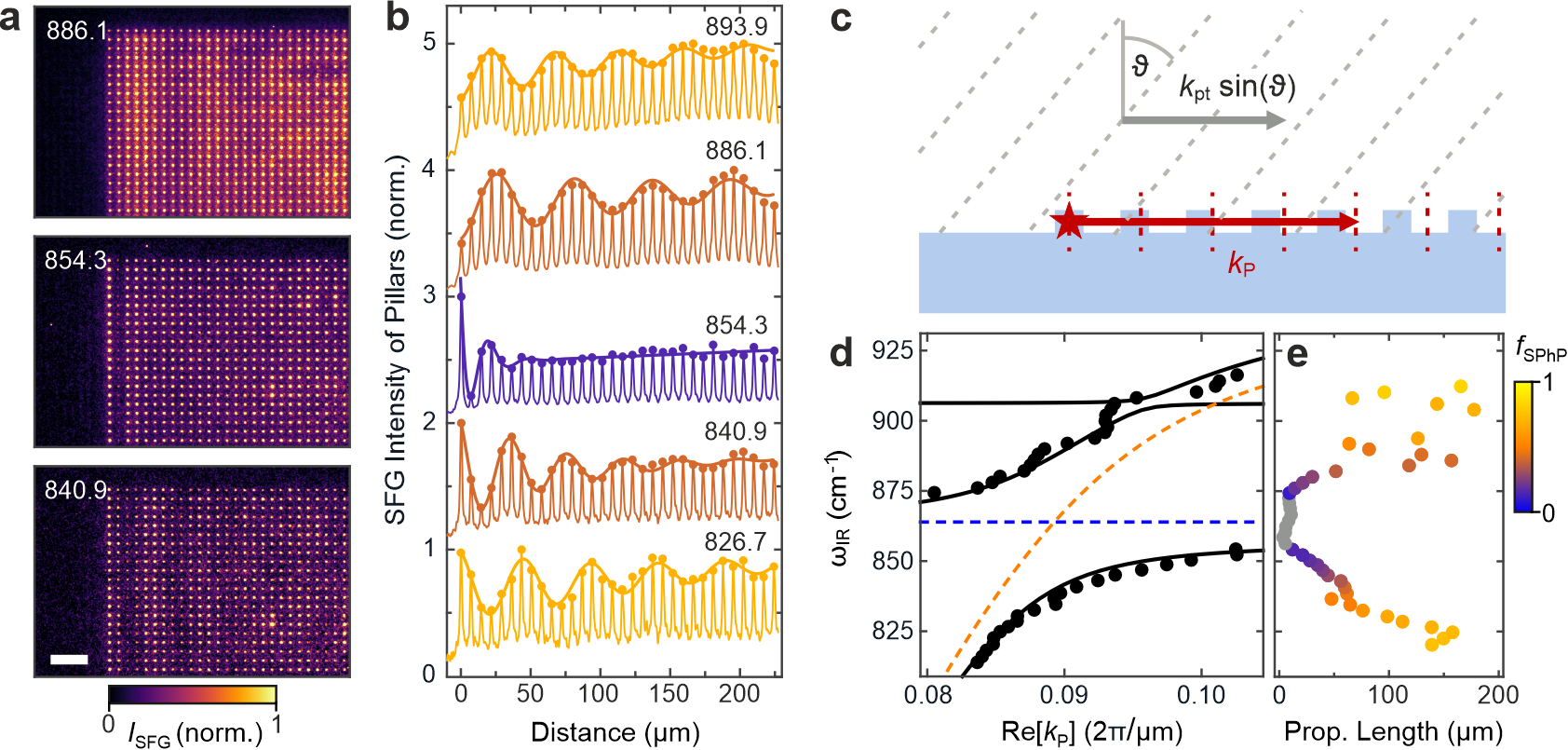}
    \caption{Interferometric imaging of polariton propagation. (a) SFG microscopy images at selected $\omega_\mathrm{IR}$ (labels in cm$^{-1}$) showing interference fringes from propagating polaritons. Scale bar 30\,µm. (b) Linescans of SFG intensity with intensity of each pillar as dots and fit with analytic model [thick lines, Eq.\,(\ref{eq:InterferometricImaging})]. (c) Sketch of edge-launched polariton (red, $k_\mathrm{P}$) and interference with illuminating IR laser ($k_\mathrm{pt} \sin{\vartheta}$) with incidence angle $\vartheta$. Dashed lines show wave fronts. (d) Polariton dispersion from interference fringes superposed on fit of data in Fig.\,2e. Data points obtained from fit of line scans with Eq.\ (\ref{eq:InterferometricImaging}) using $k_\mathrm{pt} = \omega_\mathrm{IR}/c_0$, $\vartheta = 52^\circ$, and all other parameters free. (e) Polariton propagation length $d_\mathrm{P}= 1/\mathrm{Im}[k_\mathrm{P}]$. Colors show the calculated SPhP fraction $f_\mathrm{SPhP}$ of the polaritons. }
    \label{fig:Fig3}
\end{figure*}

Analysis of the interference patterns at different IR frequencies provides us with a second independent approach to measure the polariton dispersion $\mathrm{Re}[k_\mathrm{P}(\omega)]$ (Fig.\,3d). The spatial decay away from the metasurface edge additionally allows us to measure the polariton propagation length $d_\mathrm{P}= 1/\mathrm{Im}[k_\mathrm{P}]$ (Fig.\,3e, SI Section S2). The dispersion excellently agrees with the fit of the dispersion in Fig.\,2e from resonant imaging (lines in Fig.\,3d). The two approaches are different in that interferometric imaging probes the complex polariton momentum $k_\mathrm{P}(\omega)$ for real frequencies $\omega$ set by the IR laser, while resonant imaging probes a complex polariton frequency $\omega_\mathrm{P}(k)$ through spectroscopic linewidths for real wave vectors $k$ set by the phase matching. It is remarkable that both dispersions agree so well, as complex momentum dispersions $k_\mathrm{P}(\omega)$ typically feature a backbending instead of an anti-crossing.\cite{Kusch2021, Wolff2018}

The polariton propagation length strongly changes with IR frequency from $d_\mathrm{P} > 70$\,µm to almost complete localization in the strong coupling gap (Fig.\,3e). This can be understood from the hybridization of propagating SPhPs with the localized resonances of the micropillars. The propagation length indeed scales with the calculated polariton mixing fraction $f_\mathrm{SPhP}$ that gives the proportion of propagating SPhPs in the hybridized polaritons (colors in Fig.\,3e). Polaritons with a large detuning from $\omega_\mathrm{M}$ primarily consist of propagating SPhPs and have a large propagation length accordingly. For smaller detunings the propagation length decreases and the fraction $f_\mathrm{M}$ of localized polaritons increases. In the spectrally narrow range of the lower polariton branch ($820\,$cm$^{-1} \lesssim \omega \lesssim 855\,$cm$^{-1}$) the mixing fraction varies from $f_\mathrm{SPhP} = 80\%$ to 15\%, which shows that the character of the hybridized polaritons is almost completely reversed. For slightly larger frequencies in the bandgap between the lower and upper polariton branches the polaritons become completely localized as propagation into the metasurface is forbidden (Fig.\,3e, grey).

\begin{figure*}[h!]
    \includegraphics[width=16cm]{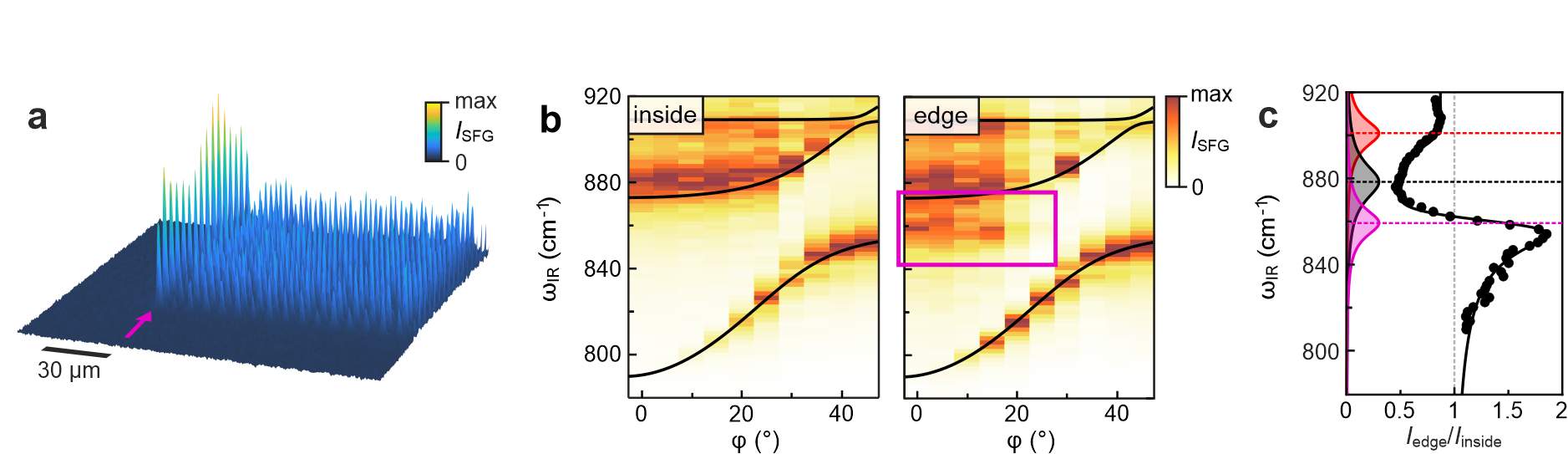}
    \caption{Polaritonic edge state in strong coupling gap.  (a) SFG microscopy image of edge state excited at $\omega_\mathrm{IR} = 854.3\ \mathrm{cm}^{-1}$. (b) Polariton dispersion from angular scan of SFG intensity evaluated at the inside of the metasurface (left, similar to Fig.\,2e) vs.\ the edge of the metasurface (right). Magenta box highlights appearance of edge state at small incidence angles $\varphi$. Lines show fit with analytic model for dispersion of polaritons across the entire array. The spectra were normalized to their maximum intensity for each angle. (c) Ratio of average SFG intensity of edge pillars $I_\mathrm{edge}$ and intensity of pillars inside the metasurface $I_\mathrm{inside}$ as function of IR frequency. Solid line is fit with analytic model with three oscillators: edge state (magenta), upper polariton (black), and dipole mode (red), see SI Section S3.}
    \label{fig:Fig4}
\end{figure*}

The localization of polaritons in the strong coupling gap gives rise to a polaritonic edge state with significantly enhanced intensity of the outermost pillar column (Fig.\,4a). The edge state appears as an additional resonance in the angular polariton dispersion from resonant imaging for small $\varphi$, when evaluating the SFG intensity only at the metasurface edge instead of the entire metasurface (Fig.\,4b, compare left and right). The edge state occurs at $\omega_\mathrm{edge} = 860\,\mathrm{cm}^{-1}$ in the polaritonic bandgap between the lower and upper polariton branches and does not shift with $\varphi$ (Fig.\,4b, magenta box). The state is visible as a selective increase of the SFG intensity at the metasurface edge (Fig.\,4a). In this spectral range, the SiC pillars in the interior of the metasurface are also weakly excited because of the proximity to the upper polariton branch at $\omega_\mathrm{P}(\varphi=0) \approx 880\,\mathrm{cm}^{-1}$. The interplay of the two resonances leads to a strong variation of the ratio $I_\mathrm{edge}/I_\mathrm{inside}$ of the intensities at the edge and inside the metasurface across the strong coupling gap (Fig.\,4c). The edge intensity resonantly transitions from an increase with respect to the metasurface interior at the lower polariton branch to a decrease at the upper polariton branch. The frequency $\omega_\mathrm{IR} \approx 855\,\mathrm{cm}^{-1}$ for the largest edge/inside intensity ratio is therefore red shifted with respect to $\omega_\mathrm{edge}$. The edge intensity also depends on the incidence angle $\varphi$ of the infrared laser. Excitation is most efficient for normal incidence ($\varphi = 0$) and vanishes for $\varphi > 25^\circ$ (Fig.\,4b), which shows that an abrupt interface with respect to the incident photons is necessary for efficient excitation of edge states.

\section{Discussion}

We introduced SFG spectro-microscopy as a technique to image phonon polaritons with highly-resolved spatial and spectral information. Our approach combines interferometric imaging, as usually employed in s-SNOM, with resonant spectral imaging, as used in nano-FTIR and reflection spectroscopy. This makes it possible to simultaneously measure the dispersion and propagation pattern of phonon polaritons. Our implementation with wide-field illumination and detection furthermore enables imaging of large sample areas at once, with an IR sub-diffractional spatial resolution that is set by the visible upconversion wavelength. The technique requires a wavelength-tunable pulsed infrared laser that is synchronized with a visible upconversion laser, as implemented in many table-top SFG spectroscopy setups,\cite{Hanninen2017, Shah2020, Khan2023} and is therefore not limited to a free-electron laser as used here. It can be applied to polaritonic materials with broken inversion symmetry such as SiC, AlN, $\alpha$-SiO$_2$, or any zincblende or wurzite-type semiconductor that have a non-zero second-order nonlinear susceptibility in the bulk. The range of materials can be potentially widened by employing inversion-broken materials as a substrate to detect the polaritonic near fields of inversion-symmetric materials.\\
\hspace*{5mm}Our approach is especially powerful for imaging metasurfaces and metacrystals, where polaritons can propagate over $>100$\,µm distances and at the same time have near fields that are confined to the sub-µm mode-volumes of the individual resonators. We demonstrated this for a millimeter-sized array of SiC micropillars where localized and propagating phonon polaritons hybridize. Spectral imaging allowed us to observe an anti-crossing from strong coupling, whereas interferometric imaging gave access to the polariton propagation length, that is strongly wavelength dependent because of mode hybridization. Making use of the IR sub-diffractional spatial resolution, we observed the localization of polaritons to the edges of the metasurface at frequencies where polariton propagation is forbidden into the metasurface. This shows that strong coupling is not just a spectral feature but has profound consequences on polariton propagation and can lead to the activation of new edge states. It will be interesting to employ SFG spectro-microscopy to image topological metasurfaces, where polaritonic or photonic states are confined to domain boundaries and can be directionally launched with circularly polarized light.\cite{Guddala2021, Ni2023, You2023} Imaging of such states is currently challenging because of the limited spatial resolution of infrared microscopes and the limited scan range of near-field techniques.\cite{Guddala2021, Sun2019} Our approach therefore extends the toolbox of imaging techniques for IR sub-diffractional photonic structures and can be applied as a fast and non-perturbative tool from the mid-IR to the terahertz spectral range.

\begin{comment}
\begin{itemize}
    \item metacrystals polaritonic/photonics
    \item edge states
    \item topological structures
    \item helical vs. non-helical
\end{itemize}
\end{comment}

\section{Methods}

\subsection{SFG spectro-microscopy}

All images and spectra were recorded with a home-built SFG microscope. The microscope was attached to the free-electron IR laser (FEL) of the Fritz-Haber-Institute Berlin.\cite{Schoellkopf2015} The FEL can be continuously tuned across a broad spectral range from the mid-IR (3000\,cm$^{-1}$) to the terahertz (200\,cm$^{-1}$) offering high-power (up to 100\,mJ), narrow-band ($\approx 0.6\%$) pulses with a macro-/micro-pulse structure. Macro pulses: 10\,Hz repetition rate, length $\tau=8$\,µs, and fluence $E=7\,$mJ/pulse in the experiments. Micro pulses: 55\,MHz repetition rate, $\tau=5\,$ps. The FEL illuminates the sample from the top at an incidence angle of $\approx 50^\circ$ with respect to the surface normal (Fig.\,1b). As a visible laser, we use a customized $\lambda_\mathrm{VIS} =532\,$nm table-top laser combining a seeder (M-PICO-LAB Nd:VAN [PR132], Montfort, 1064\,nm, 55.5\,MHz) and a flash-lamp based amplification system (Agilite 569-10, Continuum, 532\,nm, 10\,Hz, $E_{VIS}=4$\,mJ/macropulse) generating high-power, frequency-doubled laser pulses. An acousto-optic modulator between the seed laser and amplification system generates a similar macro-pulse structure as of the FEL. Temporal overlap of the micro pulses is achieved by synchronization of the seed laser with the FEL, and fine tuning with a delay stage on the optical table.\cite{kiessling2018femtosecond} The visible laser illuminates the sample from the back, parallel to the surface normal (Fig.\,1b). 
The upconverted SFG signal is collected by a long working distance objective (Mitutoyu M Plan Apo 50x, NA=0.55, working distance 13\,mm) and a 200\,mm tube lens, and detected with a gated, highly sensitive CCD camera (PI-MAX 4, Teledyne Princeton Instruments). The camera gating is synchronized with the macro pulses of the FEL and VIS lasers. Filters in front of the objective (single-band bandpass 500/24, Semrock) and in front of camera (2x tunable bandpass 547/15, Semrock) block the visible laser to only detect the SFG signal. The setup enables wide-field imaging with a field of view of 275$\times$275\,µm$^2$. Each image is averaged over 100 FEL macro pulses corresponding to an acquisition time of 10\,s/image. 

\subsection{Data acquisition and image processing}

The images were recorded using the LightField software provided with the CCD camera. An additional background image (no IR laser, only VIS illumination) was recorded and used for background correction of each image. Thereby any scattering of visible light was subtracted from the images. The background images were acquired with the same parameters as the actual SFG images, i.e., averaged over 100 VIS macro pulses.

For a full spectral scan, the FEL was automatically scanned over a frequency range from 750\,cm$^{-1}$ to 1050\,cm$^{-1}$ in 2\,cm$^{-1}$ steps, without necessity of realigning the beam path. The FEL frequency was controlled by the undulator gap size, which can be changed within a few seconds for each frequency step.\cite{Schoellkopf2015} A pyroelectric array (DIAS Infrared 128LT) coupled to a spectrometer allowed for in-situ monitoring of the FEL spectrum which was used to account for temporal frequency variations of the FEL in post-processing of the data.
%A pyroelectric detector (DIAS Infrared 128LT) allows for in-situ monitoring of the FEL spectrum which is used for spectral coorection of the IR frequencies in post-processing of the data.

\subsection{Sample fabrication}
The SiC metasurfaces (1\,mm x 1\,mm array size) were fabricated using standard lithography processes.\cite{Lu2021} In short, the 4H-SiC substrate was seeded with a thin layer of Au deposited over an adhesion layer of Cr, upon which the geometry was patterned via standard photolithography using a positive photoresist. The as-patterned wafer was then electroplated with an approximately 1\,µm thick Ni etch mask. The photoresist was then cleaned in acetone and the exposed Cr/Au seed was removed by means of an Ar ion mill plasma process. The 4H-SiC structures were then etched at a rate of about 120\,nm/min ($\sim$1\,µm etch height) in an inductively-coupled plasma (ICP) process using a combination of SF$_6$/O$_2$ chemistry optimized to yield nearly-vertical sidewall etch profiles. 

\subsection{Numerical simulations}
The polariton dispersion and local electric fields were calculated with the finite-element software COMSOL using the electromagnetic waves, frequency domain solver of the RF module. The three-dimensional unit cell of the micropillar square lattice was constructed with Floquet boundary conditions with periodicity $7.27$\,µm. SiC micropillars with diameter 1.78\,µm and height 1\,µm were placed on a semi-infinite SiC substrate. We used the anisotropic dielectric function of 4H-SiC from Ref.\,\citenum{Paarmann2016}. Perfectly matched layers and scattering boundaries were used in the vertical direction. The structure was illuminated with a plane-wave port from the top at an incidence angle $\vartheta$ with respect to the surface normal. The reflected $R$ and transmitted $T$ light were recorded with two ports and the optical absorption calculated as $A = 1-R-T$. The polariton dispersion in Fig.\ \ref{fig:Fig2}b was calculated from simulations for different incidence angles $\vartheta$ and conversion to momentum with $k(\vartheta) = 2\pi/[P(1+\sin{\vartheta})]$.

\subsection{Coupled oscillator model}
We analyzed the strong coupling with a three-mode coupled-oscillator model, as previously employed in Ref.\ \citenum{Lu2021}. The dispersion $\omega_{\mathrm{P},j}(k)$ of the three polariton branches $j$ was obtained from the eigenvalues of the coupling matrix in Eq.\ (\ref{eq:CouplingMatrix}) as $\omega_{\mathrm{P},j}(k) = \mathrm{eigval}[\mathcal{M}(k)]$. The polariton population fraction was calculated with a two-mode coupled-oscillator model for the hybridization of the monopole and SPhP modes as 
\begin{equation}
\begin{pmatrix}
 f_\mathrm{SPhP,\pm}(k)\\
 f_\mathrm{M,\pm}(k) \\
\end{pmatrix} = \mathrm{eigvec}\left[ \begin{pmatrix}
\omega_\mathrm{SPhP}(k) & g \\
g & \omega_\mathrm{M} \\
\end{pmatrix} \right]^2,
\end{equation}
with the SPhP population fraction $f_\mathrm{SPhP,\pm}$ and monopole fraction $f_\mathrm{M,\pm}$, and '$-$' for the lower polariton branch, and '$+$' for the upper branch. An analytic expression is 
\begin{equation}
    f_\mathrm{SPhP,\pm}(k) = \left[\omega_\mathrm{M}-\omega_\mathrm{SPhP}(k) \pm \sqrt{4 g^2+(\omega_\mathrm{M}-\omega_\mathrm{SPhP}(k))^2}\right]^2/4 g^2,
\end{equation} 
and $f_\mathrm{M,\pm} = 1-f_\mathrm{SPhP,\pm}$. The population fraction $f_\mathrm{SPhP,\pm}$ in Fig.\ \ref{fig:Fig3}e was calculated by assigning the data to one of the polariton branches and using the measured $k = \mathrm{Re}[k_\mathrm{P}]$ as an input parameter.

\section{Acknowledgements}
We thank Andrea Al\`u, Xiang Ni, and Simone de Liberato for fruitful discussions, and Wieland Schöllkopf and Sandy Gewinner for operating the FEL. We furthermore thank Adnan Hammud for recording the scanning electron microscopy images. G.L.\ gratefully acknowledges support from the Sir Fraser Stoddart Postdoctoral Fellowship, the Northwestern University International Institute for Nanotechnology (IIN), the Northwestern University McCormick School of Engineering, and funding from the Office of Naval Research under grant number N00014-23-1-2567. J.D.C. acknowledges funding from the Office of Naval Research under the Twist Optics MURI program (grant number N00014-22-1-2035).

\section{Competing Interests}
The authors declare no competing interests.

%\section{Supplementary Information}
%PDF: Interferometric imaging of polaritons with different propagation directions; Role of laser bandwidth for measuring the polariton propagation length; Analysis of edge/inside intensity ratio. 

%Video: Sum-frequency images recorded at different IR frequencies for $\varphi=0$. 


\begin{thebibliography}{10}
\expandafter\ifx\csname url\endcsname\relax
  \def\url#1{\texttt{#1}}\fi
\expandafter\ifx\csname urlprefix\endcsname\relax\def\urlprefix{URL }\fi
\providecommand{\bibinfo}[2]{#2}
\providecommand{\eprint}[2][]{\url{#2}}

\bibitem{caldwell2015low}
\bibinfo{author}{Caldwell, J.~D.} \emph{et~al.}
\newblock \bibinfo{title}{Low-loss, infrared and terahertz nanophotonics using
  surface phonon polaritons}.
\newblock \emph{\bibinfo{journal}{Nanophotonics}} \textbf{\bibinfo{volume}{4}},
  \bibinfo{pages}{44--68} (\bibinfo{year}{2015}).

\bibitem{Foteinopoulou2019}
\bibinfo{author}{Foteinopoulou, S.}, \bibinfo{author}{Devarapu, G. C.~R.},
  \bibinfo{author}{Subramania, G.~S.}, \bibinfo{author}{Krishna, S.} \&
  \bibinfo{author}{Wasserman, D.}
\newblock \bibinfo{title}{Phonon-polaritonics: enabling powerful capabilities
  for infrared photonics}.
\newblock \emph{\bibinfo{journal}{Nanophotonics}} \textbf{\bibinfo{volume}{8}},
  \bibinfo{pages}{2129--2175} (\bibinfo{year}{2019}).

\bibitem{Zhang2021}
\bibinfo{author}{Zhang, Q.} \emph{et~al.}
\newblock \bibinfo{title}{Interface nano-optics with van der waals polaritons}.
\newblock \emph{\bibinfo{journal}{Nature}} \textbf{\bibinfo{volume}{597}},
  \bibinfo{pages}{187--195} (\bibinfo{year}{2021}).

\bibitem{Wu2022}
\bibinfo{author}{Wu, Y.} \emph{et~al.}
\newblock \bibinfo{title}{Manipulating polaritons at the extreme scale in van
  der waals materials}.
\newblock \emph{\bibinfo{journal}{Nature Reviews Physics}}
  \textbf{\bibinfo{volume}{4}}, \bibinfo{pages}{578--594}
  (\bibinfo{year}{2022}).

\bibitem{Gubbin2022}
\bibinfo{author}{Gubbin, C.~R.}, \bibinfo{author}{De~Liberato, S.} \&
  \bibinfo{author}{Folland, T.~G.}
\newblock \bibinfo{title}{{Surface phonon polaritons for infrared
  optoelectronics}}.
\newblock \emph{\bibinfo{journal}{Journal of Applied Physics}}
  \textbf{\bibinfo{volume}{131}}, \bibinfo{pages}{030901}
  (\bibinfo{year}{2022}).

\bibitem{He2022}
\bibinfo{author}{He, M.} \emph{et~al.}
\newblock \bibinfo{title}{Anisotropy and modal hybridization in infrared
  nanophotonics using low-symmetry materials}.
\newblock \emph{\bibinfo{journal}{ACS Photonics}} \textbf{\bibinfo{volume}{9}},
  \bibinfo{pages}{1078--1095} (\bibinfo{year}{2022}).

\bibitem{Galiffi2023}
\bibinfo{author}{Galiffi, E.} \emph{et~al.}
\newblock \bibinfo{title}{{Extreme light confinement and control in
  low-symmetry phonon-polaritonic crystals}}.
\newblock \emph{\bibinfo{journal}{Accepted in Nature Reviews Materials}}
  (\bibinfo{year}{2023}).

\bibitem{li2018infrared}
\bibinfo{author}{Li, P.} \emph{et~al.}
\newblock \bibinfo{title}{Infrared hyperbolic metasurface based on
  nanostructured van der waals materials}.
\newblock \emph{\bibinfo{journal}{Science}} \textbf{\bibinfo{volume}{359}},
  \bibinfo{pages}{892--896} (\bibinfo{year}{2018}).

\bibitem{Caldwell2013}
\bibinfo{author}{Caldwell, J.~D.} \emph{et~al.}
\newblock \bibinfo{title}{Low-loss, extreme subdiffraction photon confinement
  via silicon carbide localized surface phonon polariton resonators}.
\newblock \emph{\bibinfo{journal}{Nano Letters}} \textbf{\bibinfo{volume}{13}},
  \bibinfo{pages}{3690--3697} (\bibinfo{year}{2013}).

\bibitem{razdolski2016resonant}
\bibinfo{author}{Razdolski, I.} \emph{et~al.}
\newblock \bibinfo{title}{Resonant enhancement of second-harmonic generation in
  the mid-infrared using localized surface phonon polaritons in
  subdiffractional nanostructures}.
\newblock \emph{\bibinfo{journal}{Nano letters}} \textbf{\bibinfo{volume}{16}},
  \bibinfo{pages}{6954--6959} (\bibinfo{year}{2016}).

\bibitem{Gubbin2016}
\bibinfo{author}{Gubbin, C.~R.}, \bibinfo{author}{Martini, F.},
  \bibinfo{author}{Politi, A.}, \bibinfo{author}{Maier, S.~A.} \&
  \bibinfo{author}{De~Liberato, S.}
\newblock \bibinfo{title}{Strong and coherent coupling between localized and
  propagating phonon polaritons}.
\newblock \emph{\bibinfo{journal}{Phys. Rev. Lett.}}
  \textbf{\bibinfo{volume}{116}}, \bibinfo{pages}{246402}
  (\bibinfo{year}{2016}).

\bibitem{razdolski2018second}
\bibinfo{author}{Razdolski, I.} \emph{et~al.}
\newblock \bibinfo{title}{Second harmonic generation from strongly coupled
  localized and propagating phonon-polariton modes}.
\newblock \emph{\bibinfo{journal}{Physical Review B}}
  \textbf{\bibinfo{volume}{98}}, \bibinfo{pages}{125425}
  (\bibinfo{year}{2018}).

\bibitem{Lu2021}
\bibinfo{author}{Lu, G.} \emph{et~al.}
\newblock \bibinfo{title}{Engineering the spectral and spatial dispersion of
  thermal emission via polariton–phonon strong coupling}.
\newblock \emph{\bibinfo{journal}{Nano Letters}} \textbf{\bibinfo{volume}{21}},
  \bibinfo{pages}{1831--1838} (\bibinfo{year}{2021}).

\bibitem{Hu2022}
\bibinfo{author}{Hu, X.} \emph{et~al.}
\newblock \bibinfo{title}{{Near-field nano-spectroscopy of strong mode coupling
  in phonon-polaritonic crystals}}.
\newblock \emph{\bibinfo{journal}{Applied Physics Reviews}}
  \textbf{\bibinfo{volume}{9}}, \bibinfo{pages}{021414} (\bibinfo{year}{2022}).

\bibitem{Bylinkin2021}
\bibinfo{author}{Bylinkin, A.} \emph{et~al.}
\newblock \bibinfo{title}{Real-space observation of vibrational strong coupling
  between propagating phonon polaritons and organic molecules}.
\newblock \emph{\bibinfo{journal}{Nature Photonics}}
  \textbf{\bibinfo{volume}{15}}, \bibinfo{pages}{197--202}
  (\bibinfo{year}{2021}).

\bibitem{autore2018boron}
\bibinfo{author}{Autore, M.} \emph{et~al.}
\newblock \bibinfo{title}{Boron nitride nanoresonators for phonon-enhanced
  molecular vibrational spectroscopy at the strong coupling limit}.
\newblock \emph{\bibinfo{journal}{Light: Science \& Applications}}
  \textbf{\bibinfo{volume}{7}}, \bibinfo{pages}{17172} (\bibinfo{year}{2018}).

\bibitem{Greffet2002}
\bibinfo{author}{Greffet, J.-J.} \emph{et~al.}
\newblock \bibinfo{title}{Coherent emission of light by thermal sources}.
\newblock \emph{\bibinfo{journal}{Nature}} \textbf{\bibinfo{volume}{416}},
  \bibinfo{pages}{61--64} (\bibinfo{year}{2002}).

\bibitem{Lu2022_SubArrays}
\bibinfo{author}{Lu, G.} \emph{et~al.}
\newblock \bibinfo{title}{Collective phonon–polaritonic modes in silicon
  carbide subarrays}.
\newblock \emph{\bibinfo{journal}{ACS Nano}} \textbf{\bibinfo{volume}{16}},
  \bibinfo{pages}{963--973} (\bibinfo{year}{2022}).

\bibitem{Guddala2021}
\bibinfo{author}{Guddala, S.} \emph{et~al.}
\newblock \bibinfo{title}{Topological phonon-polariton funneling in midinfrared
  metasurfaces}.
\newblock \emph{\bibinfo{journal}{Science}} \textbf{\bibinfo{volume}{374}},
  \bibinfo{pages}{225--227} (\bibinfo{year}{2021}).

\bibitem{Huber2005}
\bibinfo{author}{Huber, A.}, \bibinfo{author}{Ocelic, N.},
  \bibinfo{author}{Kazantsev, D.} \& \bibinfo{author}{Hillenbrand, R.}
\newblock \bibinfo{title}{{Near-field imaging of mid-infrared surface phonon
  polariton propagation}}.
\newblock \emph{\bibinfo{journal}{Applied Physics Letters}}
  \textbf{\bibinfo{volume}{87}}, \bibinfo{pages}{081103}
  (\bibinfo{year}{2005}).

\bibitem{Folland2019}
\bibinfo{author}{Folland, T.~G.}, \bibinfo{author}{Nordin, L.},
  \bibinfo{author}{Wasserman, D.} \& \bibinfo{author}{Caldwell, J.~D.}
\newblock \bibinfo{title}{{Probing polaritons in the mid- to far-infrared}}.
\newblock \emph{\bibinfo{journal}{Journal of Applied Physics}}
  \textbf{\bibinfo{volume}{125}}, \bibinfo{pages}{191102}
  (\bibinfo{year}{2019}).

\bibitem{alfaro2021hyperspectral}
\bibinfo{author}{Alfaro-Mozaz, F.} \emph{et~al.}
\newblock \bibinfo{title}{Hyperspectral nanoimaging of van der waals
  polaritonic crystals.}
\newblock \emph{\bibinfo{journal}{Nano Letters}} \textbf{\bibinfo{volume}{21}},
  \bibinfo{pages}{7109--7115} (\bibinfo{year}{2021}).

\bibitem{Kusch2021}
\bibinfo{author}{Kusch, P.}, \bibinfo{author}{Mueller, N.~S.},
  \bibinfo{author}{Hartmann, M.~T.} \& \bibinfo{author}{Reich, S.}
\newblock \bibinfo{title}{Strong light-matter coupling in
  ${\mathrm{mos}}_{2}$}.
\newblock \emph{\bibinfo{journal}{Phys. Rev. B}}
  \textbf{\bibinfo{volume}{103}}, \bibinfo{pages}{235409}
  (\bibinfo{year}{2021}).

\bibitem{Amarie:09}
\bibinfo{author}{Amarie, S.}, \bibinfo{author}{Ganz, T.} \&
  \bibinfo{author}{Keilmann, F.}
\newblock \bibinfo{title}{Mid-infrared near-field spectroscopy}.
\newblock \emph{\bibinfo{journal}{Opt. Express}} \textbf{\bibinfo{volume}{17}},
  \bibinfo{pages}{21794--21801} (\bibinfo{year}{2009}).

\bibitem{passler2018strong}
\bibinfo{author}{Passler, N.~C.} \emph{et~al.}
\newblock \bibinfo{title}{Strong coupling of epsilon-near-zero phonon
  polaritons in polar dielectric heterostructures}.
\newblock \emph{\bibinfo{journal}{Nano letters}} \textbf{\bibinfo{volume}{18}},
  \bibinfo{pages}{4285--4292} (\bibinfo{year}{2018}).

\bibitem{runnerstrom2018polaritonic}
\bibinfo{author}{Runnerstrom, E.~L.} \emph{et~al.}
\newblock \bibinfo{title}{Polaritonic hybrid-epsilon-near-zero modes: Beating
  the plasmonic confinement vs propagation-length trade-off with doped cadmium
  oxide bilayers}.
\newblock \emph{\bibinfo{journal}{Nano letters}} \textbf{\bibinfo{volume}{19}},
  \bibinfo{pages}{948--957} (\bibinfo{year}{2018}).

\bibitem{ni2023observation}
\bibinfo{author}{Ni, X.} \emph{et~al.}
\newblock \bibinfo{title}{Observation of directional leaky polaritons at
  anisotropic crystal interfaces}.
\newblock \emph{\bibinfo{journal}{Nature Communications}}
  \textbf{\bibinfo{volume}{14}}, \bibinfo{pages}{2845} (\bibinfo{year}{2023}).

\bibitem{Passler2022}
\bibinfo{author}{Passler, N.~C.} \emph{et~al.}
\newblock \bibinfo{title}{Hyperbolic shear polaritons in low-symmetry
  crystals}.
\newblock \emph{\bibinfo{journal}{Nature}} \textbf{\bibinfo{volume}{602}},
  \bibinfo{pages}{595--600} (\bibinfo{year}{2022}).

\bibitem{LuAPL2021}
\bibinfo{author}{Lu, G.}, \bibinfo{author}{Tadjer, M.},
  \bibinfo{author}{Caldwell, J.~D.} \& \bibinfo{author}{Folland, T.~G.}
\newblock \bibinfo{title}{{Multi-frequency coherent emission from
  superstructure thermal emitters}}.
\newblock \emph{\bibinfo{journal}{Applied Physics Letters}}
  \textbf{\bibinfo{volume}{118}}, \bibinfo{pages}{141102}
  (\bibinfo{year}{2021}).

\bibitem{Kiessling2019}
\bibinfo{author}{Kiessling, R.} \emph{et~al.}
\newblock \bibinfo{title}{Surface phonon polariton resonance imaging using
  long-wave infrared-visible sum-frequency generation microscopy}.
\newblock \emph{\bibinfo{journal}{ACS Photonics}} \textbf{\bibinfo{volume}{6}},
  \bibinfo{pages}{3017--3023} (\bibinfo{year}{2019}).

\bibitem{Niemann2022}
\bibinfo{author}{Niemann, R.} \emph{et~al.}
\newblock \bibinfo{title}{{Long-wave infrared super-resolution wide-field
  microscopy using sum-frequency generation}}.
\newblock \emph{\bibinfo{journal}{Applied Physics Letters}}
  \textbf{\bibinfo{volume}{120}}, \bibinfo{pages}{131102}
  (\bibinfo{year}{2022}).

\bibitem{barnett2022investigation}
\bibinfo{author}{Barnett, J.} \emph{et~al.}
\newblock \bibinfo{title}{Investigation of low-confinement surface phonon
  polariton launching on sic and srtio3 using scanning near-field optical
  microscopy}.
\newblock \emph{\bibinfo{journal}{Applied Physics Letters}}
  \textbf{\bibinfo{volume}{120}} (\bibinfo{year}{2022}).

\bibitem{Schoellkopf2015}
\bibinfo{author}{Sch{\"o}llkopf, W.} \emph{et~al.}
\newblock \bibinfo{title}{{The new IR and THz FEL facility at the Fritz Haber
  Institute in Berlin}}.
\newblock In \bibinfo{editor}{Biedron, S.~G.} (ed.)
  \emph{\bibinfo{booktitle}{Advances in X-ray Free-Electron Lasers
  Instrumentation III}}, vol. \bibinfo{volume}{9512}, \bibinfo{pages}{95121L}.
  \bibinfo{organization}{International Society for Optics and Photonics}
  (\bibinfo{publisher}{SPIE}, \bibinfo{year}{2015}).

\bibitem{Liu2008}
\bibinfo{author}{Liu, W.-T.} \& \bibinfo{author}{Shen, Y.~R.}
\newblock \bibinfo{title}{Sum-frequency phonon spectroscopy on
  $\ensuremath{\alpha}$-quartz}.
\newblock \emph{\bibinfo{journal}{Phys. Rev. B}} \textbf{\bibinfo{volume}{78}},
  \bibinfo{pages}{024302} (\bibinfo{year}{2008}).

\bibitem{Hanninen2017}
\bibinfo{author}{Hanninen, A.}, \bibinfo{author}{Shu, M.~W.} \&
  \bibinfo{author}{Potma, E.~O.}
\newblock \bibinfo{title}{Hyperspectral imaging with laser-scanning
  sum-frequency generation microscopy}.
\newblock \emph{\bibinfo{journal}{Biomed. Opt. Express}}
  \textbf{\bibinfo{volume}{8}}, \bibinfo{pages}{4230--4242}
  (\bibinfo{year}{2017}).

\bibitem{Shah2020}
\bibinfo{author}{Shah, S.~A.} \& \bibinfo{author}{Baldelli, S.}
\newblock \bibinfo{title}{Chemical imaging of surfaces with sum frequency
  generation vibrational spectroscopy}.
\newblock \emph{\bibinfo{journal}{Accounts of Chemical Research}}
  \textbf{\bibinfo{volume}{53}}, \bibinfo{pages}{1139--1150}
  (\bibinfo{year}{2020}).

\bibitem{Paarmann2016}
\bibinfo{author}{Paarmann, A.}, \bibinfo{author}{Razdolski, I.},
  \bibinfo{author}{Gewinner, S.}, \bibinfo{author}{Sch\"ollkopf, W.} \&
  \bibinfo{author}{Wolf, M.}
\newblock \bibinfo{title}{Effects of crystal anisotropy on optical phonon
  resonances in midinfrared second harmonic response of sic}.
\newblock \emph{\bibinfo{journal}{Phys. Rev. B}} \textbf{\bibinfo{volume}{94}},
  \bibinfo{pages}{134312} (\bibinfo{year}{2016}).

\bibitem{Wolff2018}
\bibinfo{author}{Wolff, C.}, \bibinfo{author}{Busch, K.} \&
  \bibinfo{author}{Mortensen, N.~A.}
\newblock \bibinfo{title}{Modal expansions in periodic photonic systems with
  material loss and dispersion}.
\newblock \emph{\bibinfo{journal}{Phys. Rev. B}} \textbf{\bibinfo{volume}{97}},
  \bibinfo{pages}{104203} (\bibinfo{year}{2018}).

\bibitem{Khan2023}
\bibinfo{author}{Khan, T.} \emph{et~al.}
\newblock \bibinfo{title}{Compact oblique-incidence nonlinear widefield
  microscopy with paired-pixel balanced imaging}.
\newblock \emph{\bibinfo{journal}{Opt. Express}} \textbf{\bibinfo{volume}{31}},
  \bibinfo{pages}{28792--28804} (\bibinfo{year}{2023}).

\bibitem{Ni2023}
\bibinfo{author}{Ni, X.}, \bibinfo{author}{Yves, S.}, \bibinfo{author}{Krasnok,
  A.} \& \bibinfo{author}{Alù, A.}
\newblock \bibinfo{title}{Topological metamaterials}.
\newblock \emph{\bibinfo{journal}{Chemical Reviews}}
  \textbf{\bibinfo{volume}{123}}, \bibinfo{pages}{7585--7654}
  (\bibinfo{year}{2023}).

\bibitem{You2023}
\bibinfo{author}{You, J.~W.} \emph{et~al.}
\newblock \bibinfo{title}{Topological metasurface: from passive toward active
  and beyond}.
\newblock \emph{\bibinfo{journal}{Photon. Res.}} \textbf{\bibinfo{volume}{11}},
  \bibinfo{pages}{B65--B102} (\bibinfo{year}{2023}).

\bibitem{Sun2019}
\bibinfo{author}{Sun, L.} \emph{et~al.}
\newblock \bibinfo{title}{Probing the photonic spin–orbit interactions in the
  near field of nanostructures}.
\newblock \emph{\bibinfo{journal}{Advanced Functional Materials}}
  \textbf{\bibinfo{volume}{29}}, \bibinfo{pages}{1902286}
  (\bibinfo{year}{2019}).

\bibitem{kiessling2018femtosecond}
\bibinfo{author}{Kiessling, R.} \emph{et~al.}
\newblock \bibinfo{title}{Femtosecond single-shot timing and direct observation
  of subpulse formation in an infrared free-electron laser}.
\newblock \emph{\bibinfo{journal}{Phys Rev Accel Beams}}
  \textbf{\bibinfo{volume}{21}}, \bibinfo{pages}{080702}
  (\bibinfo{year}{2018}).

\end{thebibliography}
\end{document}